\documentclass[aps,prb,twocolumn,english,showpacs,superscriptaddress,amssymb,amsfonts,amsmath]{revtex4}

\usepackage{array}
\usepackage[T1]{fontenc}
\usepackage[latin9]{inputenc}
\usepackage{amsmath}
\usepackage{dsfont}
\usepackage{amstext}

\usepackage{tocvsec2}

\usepackage{amssymb}
\usepackage{amsbsy}
\usepackage{amsthm}
\usepackage{epsfig}
\usepackage{graphicx}
\usepackage{bbm}
\usepackage{color}
\usepackage{multirow}
\usepackage[colorlinks,citecolor=blue]{hyperref}

\newcommand{\Ref}[1]{Ref.~\onlinecite{#1}}

\newcommand{\bse}{{\boldsymbol{e}}}
\newcommand{\bsg}{{\boldsymbol{g}}}

\newcommand{\ie}{{\emph{i.e.~}}}
\makeatletter

\newcommand{\Rmnum}[1]{\expandafter\@slowromancap\romannumeral #1@}
\makeatother
\newcommand{\imth}{\hspace{1pt}\mathrm{i}\hspace{1pt}}

\newcommand{\eg}{{\emph{e.g.~}}}

\newcommand{\mbz}{{\mathbb{Z}}}
\newcommand{\bea}{\begin{eqnarray}}
\newcommand{\eea}{\end{eqnarray}}
\newcommand{\bpm}{\begin{pmatrix}}
\newcommand{\epm}{\end{pmatrix}}
\newcommand{\bal}{\begin{aligned}}
\newcommand{\eal}{\end{aligned}}
\newcommand{\lra}[1]{\langle#1\rangle}

\newcommand{\be}{\begin{equation} }
\newcommand{\ee}{\end{equation} }
\newcommand{\ba}{\begin{eqnarray} }
\newcommand{\ea}{\end{eqnarray} }

\makeatother

\usepackage{babel}

\begin{document}

\title{Symmetry-induced anyon ``breeding'' in fractional quantum Hall states}

\author{Yuan-Ming Lu}
\affiliation{Department of Physics, University of California, Berkeley, CA 94720, USA}
\affiliation{Materials Science Division, Lawrence Berkeley National Laboratories, Berkeley, CA 94720}

\author{Lukasz Fidkowski}
\affiliation{Department of Physics and Astronomy, Stony Brook University, Stony
Brook, NY 11794-3800, USA.}

\date{{\small \today}}

\begin{abstract}
An exotic feature of the fractional quantum Hall effect is the emergence of anyons, which are quasiparticle excitations with fractional statistics. In the presence of a symmetry, such as $U(1)$ charge conservation, it is well known that anyons can carry fractional symmetry quantum numbers. In this work we reveal a different class of symmetry realizations: \ie anyons can ``breed'' in multiples under symmetry operation. We focus on the global Ising (${\mathbb{Z}_2}$) symmetry and show examples of these unconventional symmetry realizations in Laughlin-type fractional quantum Hall states. One remarkable consequence of such an Ising symmetry is the emergence of anyons on the Ising symmetry domain walls. We also provide a mathematical framework which generalizes this phenomenon to any Abelian topological orders. 
\end{abstract}


\maketitle

\section{Introduction}

In an electronic system where each underlying electron carries an elementary electric charge $e$, textbooks tell us that any measured electric charge should be a multiple of $e$. However, strong interactions between electrons and the associated collective electron motion can drastically alter this picture. One striking example is the discovery of the fractional quantum Hall (FQH) effect  \cite{Tsui1982}, where the low-energy quasiparticles take a fraction of an electron charge $e$ each  \cite{Laughlin1983,Goldman1995,Picciotto1997,Martin2004}. This phenomenon - collective excitations carrying symmetry quantum numbers that are fractions of those of the elementary particles - is termed \emph{fractionalization}  \cite{Laughlin1999}. In the case of Laughlin states (FQH states at filling $\nu=\frac1{2k+1},~k=1,2,\ldots$), an elementary quasiparticle can be effectively viewed as a ``fraction'' ($\frac{1}{2k+1}$) of an electron: it is neither a boson nor a fermion, but rather obeys fractional statistics\footnote{Here the statistical angle $\theta$ is defined modulo $2\pi$ through the Berry phase $e^{\imth\theta}$, obtained when two identical particles are exchanged, in two spatial dimensions.} $\theta=\frac{\pi}{2k+1}$  \cite{Arovas1984,Halperin1984} and is hence dubbed an ``anyon'' \cite{Wilczek1990B}. Therefore it is not so surprising that when global $U(1)$ symmetry associated to charge conservation is present, each anyon can have a fractional charge as well.

On the other hand, $U(1)$ charge conservation is not essential for the FQHE: Laughlin's states have been shown to be stable against arbitrary weak perturbations  \cite{Wen1990b}, which could break any symmetry. For instance we can consider Laughlin's states in systems  with a smaller symmetry group $G_s={\mathbb{Z}_2}=\{\bsg,\bsg^2=\bse\}$, \ie an Ising-type symmetry. In this case each electron has to form a representation of $G_s$, that is to say it carries an integer quantum number of the Ising symmetry. Just as in the case of $U(1)$ charge conservation symmetry, the anyons, being fractions $\frac{j}{2k+1}$ of an electron, can carry fractional charge of the Ising symmetry, \ie they can obtain a Berry phase $e^{\imth\pi j/(2k+1)}$ under the symmetry operation $\bsg$. In this work, we will expose a more exotic way in which Ising symmetry can act on Laughlin's FQH states and their cousins, the chiral spin liquids (CSLs)  \cite{Kalmeyer1987,Wen1989}.

Specifically, in the Laughlin state at $\nu=\frac1{2k+1}$, there are $2k+1$ distinct excitations, labeled $\lra a$ with $a=1,2,\cdots,2k+1$. In particular $\lra a$ can be viewed as the bound state of $a$ ``elementary'' anyons $\lra1$ (or $\frac{a}{2k+1}$ of an electron), with fractional statistics $\theta_a=\frac{a^2\pi}{2k+1}$\footnote{Notice that $\lra{2k+1}$ is an electron with statistics $\theta=\pi$, and two anyons differing by a multiple of electrons is regarded as the same: $\lra a\simeq\lra{a+\nu^{-1}}$.}.  We will construct realizations of this system that come with a natural Ising (${\mathbb{Z}_2}$) symmetry under which the elementary anyon $\lra1$ transforms into a different anyon $\lra m$ for some special $m$.  In other words, the anyons ``breed'' under the Ising symmetry: $\lra a\overset{\bsg}\longrightarrow\lra{ma}$!  Note that such a process does not conserve $U(1)$, which, as we said, is not a symmetry of our model.
A similar phenomenon also occurs in the CSL corresponding to the $\nu=\frac1{2k}$ bosonic state, which hosts $2k$ different anyons $\lra{a\mod2k}$.  We will reveal the physical basis for this exotic realization of the Ising symmetry, and analyze two explicit examples: one for Laughlin's FQH state in a 3-orbital fermion system, and the other for a CSL in a spin-1 magnet. Significant measurable consequences of this strange Ising symmetry will also be discussed.\\

\section{General discussion}
We first review statistics in ordinary realizations of the FQHE and CSL; in the next section we will write down models with the same topological order but with unconventional Ising ${\mathbb{Z}_2}$ symmetry.  For Laughlin's FQH states with $\nu=\frac1{2k+1}$ or CSLs with $\nu=\frac1{2k}$, their many-body wavefunctions on a disc geometry \cite{Laughlin1983} can be written as $\Phi_{Laughlin}\sim\prod_{i<j}(z_i-z_j)^{1/\nu}\exp[-\sum_i|z_i|^2/4]$ where $z_i=x_i+\imth y_i$ is the complex two-dimensional coordinate of $i$-th electron (in FQH states) or $\uparrow$-spins (in spin-$1/2$ CSLs). An elementary anyon $\lra1$ is created as a ``quasi-hole'' in the incompressible FQH fluid, whose wavefunction is $\Phi_{q.h.}=\prod_i(z_i-w)\Phi_{Laughlin}$ with $w$ being the coordinate of the anyon (quasi-hole).

The anyons do not breed in arbitrary multiples $m$.  Indeed, in our unconventional realization of Laughlin's FQH state, two applications of $\bsg$ must bring $\lra 1$ back to itself, {\it i.e.} ${\lra {m^2}}={\lra 1}$.  Thus $m^2-1=0$ modulo $2k+1$.  In fact, $\frac{m^2-1}{2k+1}$ must also be even.  A simple reason for this is that the Ising symmetry must preserve the fractional statistics of the anyons.  In our odd-denominator $\nu=\frac{1}{2k+1}$ FQH state, the anyon $\lra a$ has fractional statistics $\theta_a=\pi\nu a^2$ \cite{Halperin1984,Arovas1984}.  Meanwhile its mutual (braiding) statistics\footnote{The mutual statistical angle $\tilde\theta_{a,b}$ of two particles $a$ and $b$ is defined modulo $2\pi$ through the Berry phase $e^{\imth\tilde\theta_{a,b}}$, obtained by adiabatically circling particle $a$ around $b$ (or circling $b$ around $a$) once, in two spatial dimensions.} with anyon $\lra b$ is $\tilde\theta_{a,b}=2\pi\nu ab$.  All of these statistical angles are physically measurable \cite{Nayak2008} modulo $2\pi$ and should remain invariant under any symmetry operation.  In particular, $\theta_m=\theta_1$ modulo $2\pi$, so that
\bea\label{condition:laughlin} \frac{m^2-1}{2k+1} =0 \mod 2.\eea
For example, at one third filling ($\nu^{-1}=2k+1=3$), $m=5$ gives a solution to (\ref{condition:laughlin}).  An even more non-trivial solution is, for example, $m=4$ for $\nu=\frac1{15}~(k=7)$.  In this case one elementary anyon $\lra 1$ can ``give birth to a quadruplet" $\lra 4$.

For bosonic CSLs with $\nu=\frac1{2k}$, the microscopic degrees of freedom are bosonic spins rather than fermions.  The formulae for self and mutual statistics are as in the fermionic case ($\theta_a=\pi\nu a^2$ and $\tilde\theta_{a,b}=2\pi\nu ab$, but $\lra 2k$ is now a boson.  A constraint on $m$ again follows from $g$ being required to preserve topological spin:
\bea\label{condition:csl}
\lra{a}\overset{\bsg}\longrightarrow\lra{ma},~~~~~\text{if}~~\frac{{m^2-1}}{2k}=0\mod2.
\eea
There is always a nontrivial solution $m=2k-1$ to (\ref{condition:csl}). For example when $\nu=\frac14~(k=2)$, the only nontrivial solution to (\ref{condition:laughlin}) is $m=3$, which means each elementary anyon can give birth to a triplet ($\lra1\rightarrow\lra3$) under Ising symmetry. But in more complicated cases there can be other allowed multiples in which anyons can breed under the Ising symmetry. For instance, when $\nu^{-1}=2k=12$, there are 3 inequivalent nontrivial solutions to (\ref{condition:csl}): $m=5,7,11$. In the next section we will explore this example in depth, and see how to actually realize this exotic Ising symmetry in a reasonable physical system.

We want to make one more comment before delving into the examples. The filling fraction $\nu$, which equals the Hall conductance $\sigma_{xy}$ in proper units, is only well-defined for FQH states and CSLs with a $U(1)$ symmetry. When such a $U(1)$ charge/spin conservation is broken, we simply use $\nu$ to denote a FQH state/CSL which hosts $\nu^{-1}$ distinct types of anyons $\{\lra{a\mod \nu^{-1}}\}$, with fractional statistics $\theta_a$ and mutual statistics $\tilde\theta_{a,b}$. In other words, we use $\nu$ to label a gapped state which has the same topological order \cite{Wen1990b,Wen2004B} as Laughlin state \cite{Laughlin1983} $\Phi_{Laughlin}$ with filling fraction $\nu$.\\

\section{Examples}

The first example has the topological order of Laughlin's FQH state at $\nu=1/3$, and is realized in a 3-orbital fermion system. It has the following many-body wavefunction in a disc geometry \cite{Halperin1983}
\bea\label{wf:fqh/csl}
\Phi_{\nu}\big(\{z_i^{I}\}\big)=\prod_{I,J=1}^3\prod_{i<j}\big[z^{(I)}_i-z^{(J)}_j\big]^{{\bf K}_{I,J}}e^{-\sum_{i,I}\frac{|z_i^{(I)}|^2}4}.
\eea
where $z_i^{(I)}\equiv x_i^{(I)}+\imth y_i^{(I)}$ is the complex coordinate of the $i$-th fermion in the $I$-th orbital. When ${\bf K}_{I,J}<0$ the above expression actually denotes (with $\bar{z}_i^{(I)}\equiv x_i^{(I)}-\imth y_i^{(I)}$)
\bea
\big[z^{(I)}_i-z^{(J)}_j\big]^{-|{\bf K}_{I,J}|}\equiv\big[\bar{z}^{(I)}_i-\bar{z}^{(J)}_j\big]^{|{\bf K}_{I,J}|}\notag
\eea
so that $\Phi_\nu$ is well-defined. The following $3\times3$ integer matrix ${\bf K}$ in (\ref{wf:fqh/csl})
\bea
{\bf K}=\bpm-1&0&2\\0&1&0\\2&0&-1\epm\label{kmat:1/3},~~{\bf K}^{-1}=\frac13\bpm1&0&2\\0&3&0\\2&0&1\epm.
\eea
is a different representation \cite{Levin2012a,Lu2012a,sup} of Laughlin's original fermionic wavefunction for FQH states \cite{Laughlin1983}, but it supports the same set of anyon excitations and hence encodes the same topological order. To be precise, a quasihole excitation in the $I$-th orbibal can be created at coordinate $w^{(I)}$, whose wavefunction is $\Phi_{q.h.}=\prod_{i}\big[z_i^{(I)}-w^{(I)}\big]\Phi_\nu$. In such a multi-component FQH state, ${\bf K}^{-1}$ determines the statistics of anyon excitations \cite{Wen1995}. Specifically the diagonal elements of the matrix ${\bf K}^{-1}$ encode the (self) fractional statistics of quasiholes of every component in units of $\pi$, while the off-diagonal elements encode the mutual statistics between quasiholes of different components. It is easy to show \cite{sup} that a quasihole in either the 1st or 3rd orbital has fractional statistics \cite{Halperin1984} $\theta=\pi/3$, while the quasihole in the 2nd orbital is an electron with $\theta=\pi$. Also, quasiholes from the 1st orbital and 3rd orbital have nontrivial mutual statistics $\tilde\theta=2\pi/3$. Therefore a quasihole in the 1st orbital is nothing but anyon $\lra1$ in Laughlin's $\nu=1/3$ FQH state, while a quasihole in the 3rd orbital corresponds to $\lra5\simeq\lra2$. Moreover (\ref{wf:fqh/csl}) has the same Hall conductance $\sigma_{xy}=\frac13\frac{e^2}h$ as Laughlin state, if each fermion in 2nd/3rd orbital carries the same electric charge $e$ as an electron and each fermion in 1st orbital carries charge $-e$ \cite{sup}. With only $U(1)$ charge conservation, such a state is indistinguishable from the Laughlin state $\Phi_{Laughlin}$ at one third filling!

Now let us consider an Ising symmetry $\bsg$ which exchanges the 1st and 3rd orbital. From (\ref{kmat:1/3}) it is clear that the many-body wavefunction (\ref{wf:fqh/csl}) remains invariant under this Ising symmetry operation. Clearly anyon $\lra1$ becomes $\lra5$ under Ising symmetry $\bsg$: in other words, anyons breed in quintuplets ($m=5$) under Ising symmetry when $\nu=1/3$! In fact, the 3-orbital FQH state (\ref{wf:fqh/csl}) could potentially be realized by \eg fermionic cold atoms in optical lattices \cite{Hermele2009}, where the Ising symmetry of orbital exchange is present.\\

The second example is spin-1 CSLs with $\nu=\frac1{12}$. It is the simplest CSL which can support more than one pattern in which anyons breed. As mentioned earlier, there are three inequivalent nontrivial solutions to (\ref{condition:csl}): $m=5,7,11$. This is because in the $\nu=\frac1{12}$ Laughlin state, four distinct anyons $\lra1,\lra5,\lra7$ and $\lra{11}$ share the same fractional statistics $\theta=\pi/12\mod2\pi$. The many-body wavefunction for a spin-1 CSL in the $S^z$ basis still has the form (\ref{wf:fqh/csl}), where $z_i^{(I)}$ denotes the coordinate of $i$-th spin with $S^z=+1,0,-1$ for $I=1,2,3$. It turns out that several choices of the matrix ${\bf K}$ in (\ref{wf:fqh/csl}) can give states which support the set of anyon excitations of the $\nu=\frac{1}{12}$ state, and moreover different choices of such ${\bf K}$ make manifest the different possible Ising symmetries that can be realized.  Let us now explore these choices in more detail.

First of all, the matrix ${\bf K}^{(5)}$, when used in (\ref{wf:fqh/csl})
\bea
{\bf K}^{\lra5}=\bpm0&-1&3\\-1&2&-1\\3&-1&0\epm,~~~\big({\bf K}^{\lra5}\big)^{-1}=\frac1{12}\bpm1&3&5\\3&9&3\\5&3&1\epm.\label{kmat:1/12:1<->5}
\eea
describes a CSL that supports the same set of anyon excitations \cite{sup} $\{\lra{a\mod12}\}$ as the $\nu=\frac1{12}$ Laughlin state. Specifically a quasihole in the $S^z=+1$ sector corresponds to anyon $\lra1$ and a quasihole in the $S^z=-1$ sector is an anyon $\lra5$, both with fractional statistics $\theta_1=\theta_5=\frac{\pi}{12}$. Meanwhile a quasihole in $S^z=0$ sector is an anyon $\lra3$ with statistics $\theta_3=\frac{3\pi}{4}$. Under the Ising symmetry $\bsg=e^{\imth\pi\sum_{\bf r}\hat{S}^x_{\bf r}}$, which rotates spin-1's along the $x$-axis by $\pi$, anyons breed in quintuplets: $\lra1\overset{\bsg}\rightarrow\lra5$.  Note that $S^z=0$ quasiholes remain invariant under Ising symmetry since $\lra3\overset{\bsg}\rightarrow\lra{15\mod12}=\lra3$.

A different representation \cite{sup} of the $\nu=\frac1{12}$ CSL is (\ref{wf:fqh/csl}) with the following matrix ${\bf K}^{\lra7}$:
\bea
{\bf K}^{\lra7}=\bpm0&-1&2\\-1&4&-1\\2&-1&0\epm,~~~\big({\bf K}^{\lra7}\big)^{-1}=\frac1{12}\bpm1&2&7\\2&4&2\\7&2&1\epm.\label{kmat:1/12:1<->7}
\eea
Here a quasihole in the $S^z=+1$ sector is an anyon $\lra1$, an $S^z=0$ quasihole is an anyon $\lra4$ while an $S^z=-1$ quasihole corresponds to an anyon $\lra7$. Therefore under the Ising spin-flip $\bsg=e^{\imth\pi\sum_{\bf r}\hat{S}^x_{\bf r}}$, the anyons breed in septuplets: $\lra1\overset{\bsg}\rightarrow\lra7$. Again, the $S^z=0$ quasiholes $\lra4$ remain invariant under this Ising symmetry since $\lra4\overset{\bsg}\rightarrow\lra{28\mod12}=\lra4$.

\begin{figure}
 \includegraphics[width=0.35\textwidth]{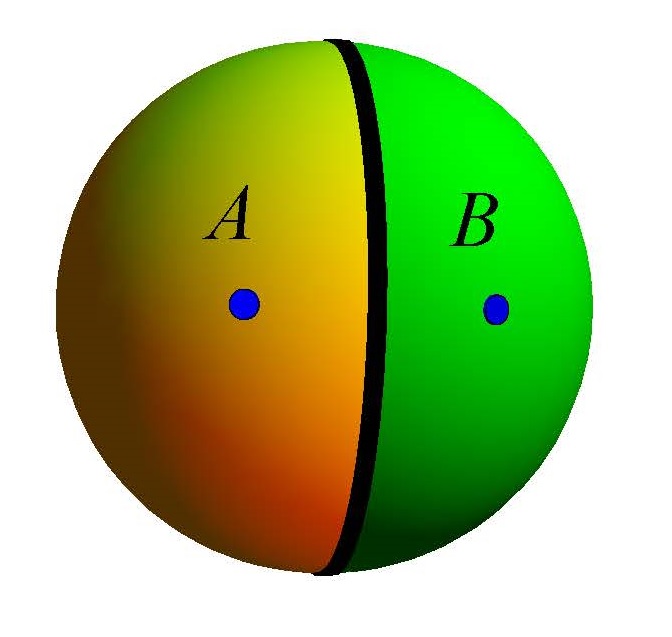}
\caption{(color online) A pair of anyons (blue dots) are created in the (unique) ground state of the FQH state (\ref{wf:fqh/csl}) on the sphere: $\lra{a}$ lies in subsystem $A$ and $\lra{-a}$ in $B$. Anyons breed in multiples of $m$ under the Ising symmetry operation. When the Ising symmetry operation is only performed on area $A$, the anyon $\lra{a}$ becomes $\lra{ma}$ and an extra anyon $\lra{a-ma}$ emerges on the boundary (the black line) separating subsystems $A$ and $B$.}\label{fig:domain wall anyon}
\end{figure}

In another representation of the $\nu=\frac1{12}$ CSL, under the Ising symmetry $\bsg=e^{\imth\pi\sum_{\bf r}\hat{S}^x_{\bf r}}$, anyons breed in multiples of $m=11$. In fact, as discussed earlier, more generally in the $\nu=\frac1{2k}$ CSL anyons can breed in multiples of $m=2k-1$. This Ising symmetry can be realized in the representation  \cite{sup}  (\ref{wf:fqh/csl}) of the $\nu=\frac1{2k}$ spin-1 CSL with the following matrix ${\bf K}$:
\bea\label{kmat:1/12:1<->11}
{\bf K}^\prime=\bpm0&1&-k\\1&0&1\\-k&1&0\epm,~\big({\bf K}^\prime\big)^{-1}=\frac1{2k}\bpm1&k&-1\\k&k^2&k\\-1&k&1\epm.
\eea
Here an $S^z=+1$ quasihole is the anyon $\lra1$, an $S^z=0$ quasihole is the anyon $\lra{k}$ and an $S^z=-1$ quasihole is the anyon $\lra{2k-1}$. The $S^z=0$ quasiholes $\lra k$ remain invariant under the Ising symmetry since $\lra{k}\overset{\bsg}\rightarrow\lra{k(2k-1)\mod2k}=\lra{k}$. These CSLs can be realized in spin-1 systems on various lattices \cite{Schroeter2007,Lu2012c}.\\

\section{Measurable consequences and domain wall anyons}

As discussed in the beginning, the existence of anyons allows symmetry to be realized and detected in an unusual way, such as fractional charges \cite{Goldman1995,Picciotto1997,Martin2004} in the presence of $U(1)$ charge conservation. In our case anyons ``breed'' in multiples, or more precisely, distinct anyons (or different superselection sectors \cite{Kitaev2006}) exchange under the Ising symmetry! This unconventional Ising symmetry has dramatic measurable effects, both in the bulk and on the boundary of the system \cite{Lu2013}.


One particular consequence of the unconventional Ising symmetry $\bsg$ is illustrated in Figure \ref{fig:domain wall anyon}.  Consider a pair of anyon excitations created from the (unique) ground state on a sphere $S$. To be specific suppose an anyon $\lra{a}$ is created in region $A$ while the other anyon $\lra{-a}\equiv\lra{-a\mod\nu^{-1}}$ lies in region $B=S\setminus A$. Now let us perform the Ising symmetry operation $\bsg$ only on subsystem $A$, so that an accompanying ``domain wall'' is created on the boundary $\partial A$ separating two regions $A$ and $B$. Since under the Ising symmetry the anyon $\lra{a}$ is transformed into $\lra{ma}$, conservation of topological charge implies that an extra anyon $\lra{a-ma}$ must be at large somewhere in the system.  In fact this new anyon $\lra{a-ma}$ stays on the domain wall $\partial A$ between the two regions, as shown in Figure \ref{fig:domain wall anyon}. The existence of domain wall anyons is a remarkable effect of such an exotic Ising symmetry.

In addition to the above phenomena, this unconventional Ising symmetry also has important measurable effects on the degenerate ground states and on the boundary excitations of the system. However in this paper we will not discuss those effects and simply refer the interested readers to \Ref{Lu2013} for further discussions.

\section{Generalizations to arbitrary Abelian topological orders} \label{sec:gen}

The ideas presented in this letter can be generalized to arbitrary Abelian topological orders, which always have $U(1)^N$ Chern-Simons ``${\bf K}$-matrix'' representations\cite{Wen1995} and can be analyzed using the theory of lattices and quadratic forms \cite{Miranda2009}.  In this section we study symmetries in bosonic Abelian topological phases from this generalized perspective.  We first describe how a lattice with an even integral symmetric bilinear form encodes the fusion and braiding structure of the anyons, and then demonstrate that the group of braided auto-equivalences - that is, permutations of the anyons that preserve fusion and braiding structure - is simply the orthogonal group of a certain ``torsion quadratic form''.  We then relate this algebraic structure to the geometric orthogonal group of an integral quadratic form, albeit one on an extended space of fields.  We start by introducing convenient notation that will facilitate our discussion.

\subsection{Lattices and Abelian Anyons}

Let ${\bf K}$ be an $N\times N$ symmetric integer matrix which is even, i.e. all the diagonal entries are even.  Consider the integer lattice ${\mathbb Z}^N \subset {\mathbb R}^N$.  ${\bf K}$ defines an inner product: if $\{ e_I \}$ is a basis for $L$ then $\langle e_I, e_J \rangle = K_{IJ}$.  We can then consider the dual lattice $L^* \subset {\mathbb R}^N$, defined to consist of those vectors whose inner product with all the vectors in $L$ is an integer.  We denote the dual basis $e_I^*\equiv\sum_JK^{-1}_{IJ}e_J$ so that $\langle e_I^*, e_J \rangle = \delta_{IJ}$.  The following notation is convenient: given $k \in {\mathbb R}^N$, we can expand

\be k = \sum_I k_I e_I^*. \ee
The column vector $(k_I)$ will be denoted ${\bf k}$.  Thus, in this notation $\langle l, k \rangle = {\bf l}^T {\bf K}^{-1} {\bf k}$.

Clearly $L \subset L^*$, so we can form $L^* / L$.  This is a finite abelian group whose elements correspond to the fractional abelian anyons in our theory.  Note that each element of $L^*/L$ has a representative in the unit cell determined by the basis $\{ e_I \}$.  That is, given $l^* \in L^*$, there exists a unique ${\overline{l^*}}$ such that $l^* - {\overline {l^*}} \in L$ and $0 \leq \langle {\overline {l^*}}, e_J^* \rangle <1$ for all $J$.  We henceforth abuse notation slightly and refer to $l^* \in L^* / L$; by this we mean that $l^* = {\overline{l^*}}$, and are actually referring to its coset in the quotient.

Since our goal is to study automorphisms of $L^* / L$ that preserve braiding and fusion rules, let us describe how to extract this data - in the form of F and R matrices, or, more formally, a unitary modular tensor category\cite{Kitaev2006} (UMTC) - from ${\bf K}$.  For completeness, we check that this data satisfies the pentagon and hexagon coherence conditions of a UMTC.  To facilitate the discussion, we introduce some more terminology.  First, consider the map $Q: L^*/L \rightarrow U(1)$ defined by

\be l^*\longrightarrow\exp\Big(\pi\imth Q(l^*)\Big) = \exp \Big( \pi\imth\langle l^*, l^* \rangle \Big)\in~U(1). \label{defQ} \ee
This map simply assigns to each anyon its topological spin.  It is well defined because $L$ is an even lattice.  Let us denote $|\det {\bf K}|$ by $|{\bf K}|$.  Each $l^* \in L^*/L$ has finite order $m$ which is a divisor of $|{\bf K}|$, \ie there is a divisor $m$ of $|{\bf K}|$ such that $m l^* \in L$.  Thus the image of $Q$ actually lies in ${\mathbb Z}_{2 |{\bf K}|} \subset U(1)$, where the inclusion is

\be r \in {\mathbb Z}_{2 |{\bf K}|} \rightarrow \exp \left(\pi\imth r / |{\bf K}|\right) \ee
We will from now on use the additive notation appropriate to ${\mathbb Z}_{2 |{\bf K}|}$ for the range of $Q$ (\ie we will be adding phases rather than multiplying their exponentials).  In this additive notation, $Q$ satisfies

\be Q(r l^*) = r^2 Q(l^*) \ee
for any $r \in {\mathbb Z}$.  Such a function is known as a `torsion quadratic form'.  In order to define the $R$ matrix, we have to somehow extend $Q$ to a `symmetric torsion bilinear form'.  This is easy when $|{\bf K}|$ is odd, but more subtle when $|{\bf K}|$ is even, so we treat the two cases separately.

When $|{\bf K}|$ is odd, then the order $m$ of any $l^* \in L^*/L$ is necessarily odd.  Therefore $\langle l^*, l^* \rangle = \langle m l^*, m l^* \rangle / m^2$ has even numerator and odd denominator, so that the image of $Q$ actually lies in ${\mathbb Z}_{|{\bf K}|} \subset {\mathbb Z}_{2 |{\bf K}|} \subset U(1)$.  Define

\be B(l^*, k^*) = \frac{|{\bf K}|+1}{2} \Big(Q(l^* + k^*) - Q(l^*) - Q(k^*) \Big) \ee
Here $B$ takes values in ${\mathbb Z}_{|{\bf K}|}$, in other words $e^{\pi\imth B(l^*,k^*)}=\exp\Big(2\pi\imth r/|{\bf K}|\Big)$ where $r\in Z_{|{\bf K}|}$.
From the definition (\ref{defQ}) of $Q$, we see that $B$ is bilinear.  It is also obviously symmetric, and satisfies $B(l^*,l^*) = Q(l^*)$.  Such a $B$ is an extension of $Q$ from a torsion quadratic form to a `symmetric torsion bilinear form'.

Returning to the construction of the $F$ and $R$ matrices, we can now set all the $F$ matrices to $1$, and let

\be R(l^*,k^*) = B(l^*,k^*) \ee
We claim that these satisfy the pentagon and hexagon equations.  Indeed, the pentagon equations are trivial, while the hexagon equations simply express the bilinearity of $B$ and its conjugate.  Also, the $S$-matrix elements are simply given up to normalization by $2B(l^*,k^*)$.  Non-degeneracy of $2B$ implies non-degeneracy of $S$.  We have thus constructed a valid MTC.

The even $|{\bf K}|$ case is more complicated, because there is now an obstruction to defining a bilinear $B$, basically because $2 \in {\mathbb Z}_{|{\bf K}|}$ is not invertible.  Indeed, we can define

\be 2B(l^*, k^*) = Q(l^* + k^*) - Q(l^*) - Q(k^*), \ee
valued in $\mathbb{Z}_{|\bf{K}|}$, which is bilinear, but for each pair $l^*, k^*$ we now have two choices of $B(l^*, k^*)$ (namely, if $2r = s$ modulo $|{\bf K}|$ then also $2r' = s$ modulo $|{\bf K}|$ with $r' = r + |{\bf K}|/2$ an integer since $|{\bf K}|$ is even).  In general it is impossible to make these choices in a way compatible with bilinearity - this is analogous, for example, to the fact that we cannot take the square root of $z$ over the entire complex plane in a way compatible with continuity.

Nevertheless, the failure of bilinearity can be compensated by appropriate signs in the F-matrix in such a way that the entire structure obeys the pentagon and hexagon equations.  We describe this construction in detail in Appendix \ref{sec:appA}.

\subsection{Braided Auto-equivalences}

Now that we have shown how to extract the UMTC structure from ${\bf K}$, we study which permutations of the anyons are compatible with this UMTC structure.  Potential symmetry actions will be restricted to permutations of this form.  Let us first make some general remarks that apply regardless of the parity of $|{\bf K}|$.  First, in order to preserve fusion rules, a putative braided equivalence $g$ must be a group automorphism of the group $L^*/L$.  Second, $g$ must preserve topological spins, since these are topological invariants, i.e. $g$ must preserve the torsion quadratic form $Q$ on $L^*/L$.  Now, there is a theorem (Proposition 9.4 in \Ref{Miranda2009}) that any such automorphism $g$ of $L^*/L$ lifts to a stable isomorphism from $L$ and $L \oplus M$, where $M$ is a unimodular lattice - in fact, $M$ can be taken to be a direct sum of hyperbolic planes\cite{Miranda2009}.

Unwound, this means the following.  Form the extension ${\bf K}_e$ of ${\bf K}$ to $N+2N'$ dimensions by adding a direct sum $N'$ copies of the two by two matrix

\be {\bf H}=\left( \begin{array}{cc} 0&1\\1&0 \end{array} \right), \ee

\be {\bf K}_e = {\bf K} \oplus {\bf H}^{\otimes N'} \ee
where $N'$ is a positive integer.  The extended lattice corresponding to ${\bf K}_e$ is simply $L_e=L \oplus M$, while its dual $L_e^* = L^* \oplus M$ since $M$ is unimodular.  Hence the quotient $L_e^*/L_e$ can be canonically identified with $L^*/L$.  Now, suppose ${\bf W}$ is a $N+2N'$ by $N+2N'$ unimodular integer matrix (i.e. ${\bf W}$ has integer entries and determinant $1$) which satisfies:

\be {\bf W}^{T} {\bf K}_e {\bf W} = {\bf K}_e. \ee
Then ${\bf W}$ defines an isometry which maps the lattices $L_e, L_e^*$ to themselves, and hence defines an automorphism $g_W$ of $L_e^*/L_e = L^*/L$.  Now the claim is that given {\it any} group automorphism $g: L^*/L \rightarrow L^*/L$ preserving the torsion quadratic form $Q$, there exists some $N'$ and $W$ constructed as above for which $g = g_W$.  Thus, the abstract algebraic objects we started with, namely automorphisms of the torsion quadratic form $Q$, have a geometrical interpretation as isometries, provided we add enough extra `trivial' dimensions.  Note that adding extra dimensions may already be necessary even with one by one K matrices.  For example, if ${\bf K} = 12$ as in our example above, then $x \rightarrow 5x \mod 12$ is an automorphism of $Q$ which can not be represented geometrically in one dimension.  However, it can be represented geometrically once one adds one copy of the hyperbolic plane, so that

\be {\bf K}_e = \left(\begin{array}{ccc} 0&1&0\\1&0&0\\0&0&12 \end{array} \right). \ee
Indeed, in this case
\be {\bf W} = \left(\begin{array}{ccc} 3&-2&12\\-2&3&-12\\1&-1&5 \end{array} \right) \ee
does the job.

We have shown that any putative braided auto-equivalence of the UMTC associated with $L^*/L$ must at least be an automorphism of the torsion quadratic for $Q$ defining the topological spins.  Now we state a converse: any automorphism of the torsion quadratic form actually defines a valid braided auto-equivalence.  When $|{\bf K}|$ is odd this is quite easy to show, because any automorphism of the torsion quadratic form automatically preserves the bilinear form $B(l^*,k^*)$, and hence the R matrices.  Since the F matrices are all equal to $1$, $g$ trivially preserves them, and thus defines a braided automorphism of $L^*/L$.  When $|{\bf K}|$ is even, the argument is more subtle, but the conclusion remains the same.

Thus we have completely characterized the braided auto-equivalences of a bosonic theory with K-matrix ${\bf K}$ - they are just the automorphisms of its torsion quadratic form.  Furthermore, we have shown that any such automorphism can be lifted to an isometry of a stably equivalent K-matrix, generalizing the examples given in the beginning of this paper.

\section{Discussion}

In this work we presented a class of exotic realizations of global Ising symmetry in FQH states that host the same set of anyon excitations $\{\lra{a\mod\nu^{-1}}\}$ as Laughlin states with filling fraction $\nu=\frac1{2k+1}$ or $\nu=\frac1{2k}$. Under the Ising symmetry operation, anyons can ``breed'' in multiples of $m$, \ie $\lra{a}\overset{\bsg}\rightarrow\lra{ma}$. Examples in $\nu=\frac1{2k+1}$ FQH state and $\nu=\frac1{2k}$ chiral spin liquids were demonstrated. We also discussed significant measurable consequences of this Ising symmetry, \eg the fact that anyons can emerge on the ``domain wall'' of the Ising symmetry. These symmetry-enriched FQH states may potentially be realized in fermions in optical lattices or in spin-1 magnets. Such a phenomenon, \ie anyons exchanging or breeding under symmetry operations, can be generalized to an arbitrary Abelian topological order.

After the completion of this work we became aware of \Ref{Cano2013} in which similar ideas are introduced.

\section{Acknowledgements}

This work is supported by Office of BES, Materials Sciences Division of the U.S. DOE under contract No. DE-AC02-05CH11231 (YML).  We are grateful for useful discussions with A. Kitaev, N. Lindner, and A. Vishwanath.

\appendix

\section{Why are Laughlin states $\Phi_{Laughlin}$ equivalent to 3-component states $\Phi_\nu$?}

In this section we explain in detail why Laughlin's states\cite{Laughlin1983} at filling fraction $\nu=\frac1{2k+1}$ (or $\nu=\frac1{2k}$)
\bea\label{wf:laughlin}
\Phi_{Laughlin}\big(\{z_i\}\big)=\prod_{i<j}(z_i-z_j)^{1/\nu}\exp[-\sum_i|z_i|^2/4].
\eea
are equivalent to the following 3-component fractional quantum Hall (FQH) states\cite{Halperin1983}
\bea\label{wf:3-component}
\Phi_{\nu}\big(\{z_i^{I}\}\big)=\prod_{I,J=1}^3\prod_{i<j}\big[z^{(I)}_i-z^{(J)}_j\big]^{{\bf K}_{I,J}}e^{-\sum_{i,I}\frac{|z_i^{(I)}|^2}4}.
\eea
with proper choices of symmetric integer matrix ${\bf K}$. By ``equivalent'' we mean the two gapped states (\ref{wf:laughlin}) and (\ref{wf:3-component}) supports the same set of anyon excitations\cite{Halperin1984,Arovas1984,Wilczek1990B}. In other words they share the same topological order\cite{Wen1995,Wen2004B}.

First of all, a fermionic Laughlin state (\ref{wf:laughlin}) with $\nu=\frac1{2k+1}$ (so that many-body wavefunction is anti-symmetric) is equivalent to (\ref{wf:3-component}) with
\bea\label{kmat:laughlin:1/2k+1:0}
{\bf K}_0=\bpm2k+1&&\\&1&\\&&-1\epm
\eea
which\cite{Levin2012a,Lu2012a} simply stacks a layer of integer quantum Hall (IQH) state of Hall conductance $\sigma_{xy}=1$ (in unit of $e^2/h$) and another layer of IQH state of $\sigma_{xy}=-1$ on top of the Laughlin state. This is because an IQH state of fermions only hosts fermionic excitations: it doesn't bring in anyon quasiparticles into $\nu=\frac1{2k+1}$ Laughlin's FQH state. And the total Hall conductance $\sigma_{xy}=\frac1{2k+1}$ is the same after stacking the two layers. In fact these two states, \ie (\ref{wf:laughlin}) with $\nu=\frac1{2k+1}$ and (\ref{wf:3-component}) with (\ref{kmat:laughlin:1/2k+1:0}) are indistinguishable in the presence only $U(1)$ charge conservation.

Now notice that in wavefunction (\ref{wf:3-component}) one can choose how the three components ($I=1,2,3$) of fermions are defined. For examples with $\nu=\frac13~(k=1)$, when we choose matrix (\ref{kmat:laughlin:1/2k+1:0}) in wavefunction (\ref{wf:3-component}) we also choose the definition of fermions for the 3-components: $f_{1,2,3}$. We can redefine electron operators in a different basis: \eg define the new 3-component fermions as $f_{2,3}^\prime=f_{2,3}$ but $f_1^\prime\sim f_1(f_3^\dagger)^2$ (\ie combine two holes of 3rd component with one fermion of 1st component to form the new fermion of 1st component). In this new basis the matrix becomes
\bea
&\notag{\bf K}={\bf X}^T{\bf K}_0{\bf X}=\bpm-1&0&2\\0&1&0\\2&0&-1\epm,\\
&{\bf X}=\bpm1&0&0\\0&1&0\\-2&0&1\epm,~~~\det{\bf X}=1.
\eea
in many-body wavefunction (\ref{wf:3-component}). In general such a change of basis (and redefinition of ``electrons'') is implemented\cite{Read1990,Frohlich1991,Wen1992,Lu2012a} by a $N\times N$ unimodular matrix ${\bf X}\in GL(N,\mbz)$~(note $\det{\bf X}=\pm1$) for a symmetric $N\times N$ integer matrix ${\bf K}$. The $U(1)$ charge conservation can be labeled by an integer \emph{charge vector}\cite{Wen1992} ${\bf q}_0=(1,1,1)^T$: which means each fermion of any component carries unit $U(1)$ charge. Under a change of basis implemented by $GL(N,\mbz)$ transformation ${\bf X}$ the charge vector ${\bf q}$ changes as
\bea
{\bf q}_0=(1,1,1)^T\rightarrow{\bf q}={\bf X}^T{\bf q}_0=(-1,1,1)^T.
\eea
This means each new fermion of 1st component carries $U(1)$ charge $-1$, while each fermion of 2nd or 3rd component carries unit $U(1)$ charge. Therefore we've shown that matrix ${\bf K}$ with charge vector ${\bf q}$ in (\ref{wf:3-component}) represents the same FQH state as matrix ${\bf K}_0$ with charge vector ${\bf q}_0$ in (\ref{wf:3-component}), in the presence of $U(1)$ symmetry. Both representations are indistinguishable with the original Laughlin state (\ref{wf:laughlin}) with $\nu=1/3$.\\

Now let's turn to Laughlin's chiral spin liquids (CSLs), \ie $\nu=\frac1{2k}$ in (\ref{wf:laughlin}), where the microscopic degrees of freedom are bosonic spins instead of fermions. Like in the fermion case, a two-dimensional trivial insulator of bosons can be stacked on top of the CSL and it doesn't change the anyon excitations of the system at all. Such a trivial boson insulator is given by the lower $2\times2$ part\cite{Levin2012a,Lu2012a} of the following matrix
\bea\label{kmat:1/2k laughlin}
{\bf K}_b=\bpm2k&0&0\\0&1&0\\0&0&1\epm
\eea
In other words (\ref{wf:3-component}) with this matrix ${\bf K}_0$ and original Laughlin state (\ref{wf:laughlin}) with $\nu=\frac1{2k}$ are indistinguishable in the absence of extra symmetries. Again one can change the basis by $GL(N,\mbz)$ transformations ${\bf X}$, which yield
\bea\label{kmat:1/2k:1<->2k-1}
{\bf K}^\prime={\bf X}_b^T{\bf K}_b{\bf X}_b=\bpm0&1&-k\\1&0&1\\-k&1&0\epm,~~~{\bf X}_b=\bpm1&0&0\\-k&1&0\\1&0&1\epm.
\eea

Moreover when $\nu=\frac1{12}$ \ie $k=6$, (\ref{kmat:1/2k laughlin}) has the following two different representations
\bea\label{kmat:1/12:1<->5}
&{\bf K}^{\lra5}={\bf X}_1^T{\bf K}_b{\bf X}_1=\bpm0&-1&3\\-1&2&-1\\3&-1&0\epm,\\
&\notag{\bf X}_1=\bpm1&0&0\\-3&1&0\\2&1&-1\epm\in GL(3,\mbz).
\eea
and
\bea\label{kmat:1/12:1<->7}
&{\bf K}^{\lra7}={\bf X}_2^T{\bf K}_b{\bf X}_2=\bpm0&-1&2\\-1&4&-1\\2&-1&0\epm,\\
&\notag{\bf X}_2=\bpm1&0&0\\-2&1&0\\3&2&-1\epm\in GL(3,\mbz).
\eea
They are all indistinguishable with the original Laughlin state (\ref{wf:laughlin}) with $\nu=\frac1{12}$.

\section{Fusion and Braiding rules for even $|{\bf K}|$} \label{sec:appA}

Here we give the technical construction of the F and R symbols for the case of an abelian theory defined by a ${\bf K}$ matrix with even $|{\bf K}|$.  The conventions are as in section \ref{sec:gen}.  We start by defining the F symbols.  First of all, as an abstract abelian group $L^*/L$ splits as a direct product of its Sylow $p$-subgroups over prime divisors $p$ of $|{\bf K}|$:

\be L^*/L = \prod_{p} A_p = A_2 \times A_{\rm{odd}} \ee
where

\be A_{\rm{odd}} = \prod_{p\geq 3} A_p. \label{2split} \ee
Here $A_p$ is defined as the set of all $l^* \in L^*/L$ whose order is a power of $p$.  Given $l^* \in L^*/L$, let us denote its decomposition with respect to (\ref{2split}) by $(l^*_2, l^*_{\rm{odd}})$.  For $l^*, k^* \in L^*/L$, we note that $2B(l^*_2, k^*_{\rm{odd}})$ is killed when multiplied by a power of $2$ (using bilinearity in the first variable together with the fact that the order of $l^*_2$ is a power of $2$), and is also killed when multiplied by an odd number (using bilinearity in the second variable, whose order is odd).  By Euclid's algorithm there is a linear combination of these equal to $1$, which means that $2B(l^*_2, k^*_{\rm{odd}})$ is itself $0$.  By symmetry $2B(k^*_2, l^*_{\rm{odd}})$ also vanishes, and we obtain

\be 2B(l^*, k^*) = 2B(l^*_2, k^*_2) + 2B(l^*_{\rm{odd}}, k^*_{\rm{odd}}) \label{orth}\ee
Thus $A_2$ and $A_{\rm{odd}}$ are orthogonal under $2B$.  We now define $B$ on $A_2$ and $A_{\rm{odd}}$ separately.  On $A_{\rm{odd}}$ we can proceed as in the previous section, since $|A_{\rm{odd}}|$ is odd and division by $2$ is not a problem.  For $A_2$, we proceed as follows.  Given a value of $2B(l^*_2, k^*_2) \in {\mathbb Z}_{|\bf{K}|} = j$, we set $B(l^*_2, k^*_2) = j \in {\mathbb Z}_{2|\bf{K}|}$.  Note that this is effectively a division by $2$ if we look at ${\mathbb Z}_{|\bf{K}|} \subset {\mathbb Z}_{2|\bf{K}|} \subset U(1)$, and that it is {\it not} a linear map ${\mathbb Z}_{|\bf{K}|} \rightarrow {\mathbb Z}_{2|\bf{K}|}$.

We can now define
\be B(l^*, k^*) = B(l^*_2, k^*_2) + B(l^*_{\rm{odd}}, k^*_{\rm{odd}}) \label{decompB} \ee
on all of $L^*/L$.  In order for this to be a useful definition, we should at least make sure that $B(l^*, l^*) = Q(l^*)$, since the latter is already defined on all of $L^*/L$.  This follows from the fact that any torsion quadratic form (in particular $Q$) respects the splitting into Sylow subgroups.  The proof of this fact is slightly more involved than that of its analogue for torsion bilinear forms (\ref{orth}) - the $2$-Sylow subgroup must be treated carefully (see \Ref{Miranda2009}, Section II.5)  $B$, as defined in (\ref{decompB}), only fails to be bilinear on $A_2$; the second term on the right hand side of (\ref{decompB}) is bilinear by the arguments of the previous section since $A_{\rm odd}$ has odd cardinality.

We now prove a technical result about $A_2$ which will be important later.  We view $L \subset L^*$ as embedded in ${\mathbb R}^N$.  We claim that there exists a vector subspace $S \subset {\mathbb R}^N$ with the following properties: 1) any vector $l \in L \cap S$ has even inner product with any other $k \in L$ and 2) $(L^* \cap S) / (L \cap S) \subset L^*/L$ contains $A_2$.  To see this, chose a minimal set of elements $e_1, \ldots, e_r \in L^*$ which generate $A_2$ in $L^*/L$, and let $S$ be their span over ${\mathbb R}$.  Then property 2 is automatic.  Furthermore if $l \in L \cap S$, then we can write $l = n_1 e_1 + \ldots + n_r e_r$ with integers $n_i$.  We claim that $l/2 \in L^* \cap S$.  For if not, then at least one of the $n_i$ is odd, and since the order of $e_i$ is a power of $2$ and hence relatively prime to $n_i$, by Euclid's algorithm we can express $e_i$ as a linear combination of the other $e_j$ modulo $L$, violating the assumed minimality of the generating set $e_1, \ldots, e_r$.  Thus $l/2$ has integer inner products with all $k\in L$, meaning that $l$ has even inner products with all $k\in L$.  We will from now on choose our unit cell of $L \in {\mathbb R}^N$ in a way that is compatible with the choice of $S$ - specifically, we ensure that all the representatives of $A_2$ in our chosen unit cell also lie in $S$.  This will become important in checking the pentagon equation for the F symbols we now construct.

Define, for $l^*, j^*, k^* \in L^*/L$,

\be F_{l^*, j^*, k^*} = \exp \left(\pi i \langle l^*, \overline{j^*_2} + \overline{k^*_2} - \overline{j^*_2+k^*_2} \rangle \right) \label{Fmatrix} \ee
Note that the $F$-symbols defined in (\ref{Fmatrix}) are in fact signs.  To check that they satisfy the pentagon equation, we simply have to show that the $3$-cochain defined by (\ref{Fmatrix}) is actually a $3$-cocycle.  We explicitly calculate

\begin{align} &dF(l^*, j^*, k^*,m^*) = \\ &\exp \left( \pi i \langle \overline{l^*_2} + \overline{j^*_2} - \overline{l^*_2 + j^*_2}, \overline{k^*_2} + \overline{m^*_2} - \overline{k^*_2 + m^*_2} \rangle \right) \end{align}
In order for this to vanish, we have to check that the inner product above is an ${\it even}$ integer.  But both of the vectors entering the inner product are in $L\cap S$, and hence the inner product is even by the technical result above.
Thus the $F$-symbol satisfies the pentagon identity, and defines a cohomology class

\be [F] \in H^3(L^*/L, {\mathbb Z}_2). \ee
We can also look at the cohomology class of $F$ in $H^3(L^*/L, U(1))$.

Now we have to check the hexagon equation.  Let us consider a hexagon diagram which one either fuses $j^*$ and $k^*$ and braids the product around $l^*$, or braids $j^*$ and $k^*$ individually and then fuses.  This diagram contains three phases that come from $R$-matrix moves and three signs that come from $F$-matrix moves.  We can check directly that the (appropriately ordered) product of the $R$-matrix phases is

\be \exp \left(\pi i \langle l_2^*, \overline{j_2^*} + \overline{k_2^*} - \overline{j_2^*+k_2^*} \rangle \right. \ee
Meanwhile, the $F$-matrix phases multiply out to

\be F_{l^*,j^*,k^*} F_{j^*,l^*,k^*} F_{j^*,k^*,l^*} = F_{l^*,j^*,k^*} \ee
since $F_{j^*,l^*,k^*} = F_{j^*, k^*, l^*}$.
The $F$ and $R$ matrix phases cancel using (\ref{Fmatrix}), so that the hexagon equation is satisfied.  The other hexagon equation is simply the complex conjugate of this.

%

\begin{thebibliography}{50}
\expandafter\ifx\csname natexlab\endcsname\relax\def\natexlab#1{#1}\fi
\expandafter\ifx\csname bibnamefont\endcsname\relax
  \def\bibnamefont#1{#1}\fi
\expandafter\ifx\csname bibfnamefont\endcsname\relax
  \def\bibfnamefont#1{#1}\fi
\expandafter\ifx\csname citenamefont\endcsname\relax
  \def\citenamefont#1{#1}\fi
\expandafter\ifx\csname url\endcsname\relax
  \def\url#1{\texttt{#1}}\fi
\expandafter\ifx\csname urlprefix\endcsname\relax\def\urlprefix{URL }\fi
\providecommand{\bibinfo}[2]{#2}
\providecommand{\eprint}[2][]{\url{#2}}

\bibitem[{\citenamefont{Tsui et~al.}(1982)\citenamefont{Tsui, Stormer, and
  Gossard}}]{Tsui1982}
\bibinfo{author}{\bibfnamefont{D.~C.} \bibnamefont{Tsui}},
  \bibinfo{author}{\bibfnamefont{H.~L.} \bibnamefont{Stormer}},
  \bibnamefont{and} \bibinfo{author}{\bibfnamefont{A.~C.}
  \bibnamefont{Gossard}}, \bibinfo{journal}{Phys. Rev. Lett.}
  \textbf{\bibinfo{volume}{48}}, \bibinfo{pages}{1559} (\bibinfo{year}{1982}).

\bibitem[{\citenamefont{Laughlin}(1983)}]{Laughlin1983}
\bibinfo{author}{\bibfnamefont{R.~B.} \bibnamefont{Laughlin}},
  \bibinfo{journal}{Phys. Rev. Lett.} \textbf{\bibinfo{volume}{50}},
  \bibinfo{pages}{1395} (\bibinfo{year}{1983}).

\bibitem[{\citenamefont{Goldman and Su}(1995)}]{Goldman1995}
\bibinfo{author}{\bibfnamefont{V.~J.} \bibnamefont{Goldman}} \bibnamefont{and}
  \bibinfo{author}{\bibfnamefont{B.}~\bibnamefont{Su}},
  \bibinfo{journal}{Science} \textbf{\bibinfo{volume}{267}},
  \bibinfo{pages}{1010} (\bibinfo{year}{1995}).

\bibitem[{\citenamefont{de~Picciotto et~al.}(1997)\citenamefont{de~Picciotto,
  Reznikov, Heiblum, Umansky, Bunin, and Mahalu}}]{Picciotto1997}
\bibinfo{author}{\bibfnamefont{R.}~\bibnamefont{de~Picciotto}},
  \bibinfo{author}{\bibfnamefont{M.}~\bibnamefont{Reznikov}},
  \bibinfo{author}{\bibfnamefont{M.}~\bibnamefont{Heiblum}},
  \bibinfo{author}{\bibfnamefont{V.}~\bibnamefont{Umansky}},
  \bibinfo{author}{\bibfnamefont{G.}~\bibnamefont{Bunin}}, \bibnamefont{and}
  \bibinfo{author}{\bibfnamefont{D.}~\bibnamefont{Mahalu}},
  \bibinfo{journal}{Nature} \textbf{\bibinfo{volume}{389}},
  \bibinfo{pages}{162} (\bibinfo{year}{1997}), ISSN \bibinfo{issn}{0028-0836}.

\bibitem[{\citenamefont{Martin et~al.}(2004)\citenamefont{Martin, Ilani,
  Verdene, Smet, Umansky, Mahalu, Schuh, Abstreiter, and Yacoby}}]{Martin2004}
\bibinfo{author}{\bibfnamefont{J.}~\bibnamefont{Martin}},
  \bibinfo{author}{\bibfnamefont{S.}~\bibnamefont{Ilani}},
  \bibinfo{author}{\bibfnamefont{B.}~\bibnamefont{Verdene}},
  \bibinfo{author}{\bibfnamefont{J.}~\bibnamefont{Smet}},
  \bibinfo{author}{\bibfnamefont{V.}~\bibnamefont{Umansky}},
  \bibinfo{author}{\bibfnamefont{D.}~\bibnamefont{Mahalu}},
  \bibinfo{author}{\bibfnamefont{D.}~\bibnamefont{Schuh}},
  \bibinfo{author}{\bibfnamefont{G.}~\bibnamefont{Abstreiter}},
  \bibnamefont{and} \bibinfo{author}{\bibfnamefont{A.}~\bibnamefont{Yacoby}},
  \bibinfo{journal}{Science} \textbf{\bibinfo{volume}{305}},
  \bibinfo{pages}{980} (\bibinfo{year}{2004}).

\bibitem[{\citenamefont{Laughlin}(1999)}]{Laughlin1999}
\bibinfo{author}{\bibfnamefont{R.~B.} \bibnamefont{Laughlin}},
  \bibinfo{journal}{Rev. Mod. Phys.} \textbf{\bibinfo{volume}{71}},
  \bibinfo{pages}{863} (\bibinfo{year}{1999}).

\bibitem[{\citenamefont{Arovas et~al.}(1984)\citenamefont{Arovas, Schrieffer,
  and Wilczek}}]{Arovas1984}
\bibinfo{author}{\bibfnamefont{D.}~\bibnamefont{Arovas}},
  \bibinfo{author}{\bibfnamefont{J.~R.} \bibnamefont{Schrieffer}},
  \bibnamefont{and} \bibinfo{author}{\bibfnamefont{F.}~\bibnamefont{Wilczek}},
  \bibinfo{journal}{Phys. Rev. Lett.} \textbf{\bibinfo{volume}{53}},
  \bibinfo{pages}{722} (\bibinfo{year}{1984}).

\bibitem[{\citenamefont{Halperin}(1984)}]{Halperin1984}
\bibinfo{author}{\bibfnamefont{B.~I.} \bibnamefont{Halperin}},
  \bibinfo{journal}{Phys. Rev. Lett.} \textbf{\bibinfo{volume}{52}},
  \bibinfo{pages}{1583} (\bibinfo{year}{1984}).

\bibitem[{\citenamefont{Wilczek}(1990)}]{Wilczek1990B}
\bibinfo{author}{\bibfnamefont{F.}~\bibnamefont{Wilczek}},
  \emph{\bibinfo{title}{Fractional Statistics and Anyon Superconductivity}}
  (\bibinfo{publisher}{World Scientific Pub Co Inc}, \bibinfo{year}{1990}).

\bibitem[{\citenamefont{Wen and Niu}(1990)}]{Wen1990b}
\bibinfo{author}{\bibfnamefont{X.~G.} \bibnamefont{Wen}} \bibnamefont{and}
  \bibinfo{author}{\bibfnamefont{Q.}~\bibnamefont{Niu}},
  \bibinfo{journal}{Phys. Rev. B} \textbf{\bibinfo{volume}{41}},
  \bibinfo{pages}{9377} (\bibinfo{year}{1990}).

\bibitem[{\citenamefont{Kalmeyer and Laughlin}(1987)}]{Kalmeyer1987}
\bibinfo{author}{\bibfnamefont{V.}~\bibnamefont{Kalmeyer}} \bibnamefont{and}
  \bibinfo{author}{\bibfnamefont{R.~B.} \bibnamefont{Laughlin}},
  \bibinfo{journal}{Phys. Rev. Lett.} \textbf{\bibinfo{volume}{59}},
  \bibinfo{pages}{2095} (\bibinfo{year}{1987}).

\bibitem[{\citenamefont{Wen et~al.}(1989)\citenamefont{Wen, Wilczek, and
  Zee}}]{Wen1989}
\bibinfo{author}{\bibfnamefont{X.~G.} \bibnamefont{Wen}},
  \bibinfo{author}{\bibfnamefont{F.}~\bibnamefont{Wilczek}}, \bibnamefont{and}
  \bibinfo{author}{\bibfnamefont{A.}~\bibnamefont{Zee}},
  \bibinfo{journal}{Phys. Rev. B} \textbf{\bibinfo{volume}{39}},
  \bibinfo{pages}{11413} (\bibinfo{year}{1989}).

\bibitem[{\citenamefont{Nayak et~al.}(2008)\citenamefont{Nayak, Simon, Stern,
  Freedman, and Das~Sarma}}]{Nayak2008}
\bibinfo{author}{\bibfnamefont{C.}~\bibnamefont{Nayak}},
  \bibinfo{author}{\bibfnamefont{S.~H.} \bibnamefont{Simon}},
  \bibinfo{author}{\bibfnamefont{A.}~\bibnamefont{Stern}},
  \bibinfo{author}{\bibfnamefont{M.}~\bibnamefont{Freedman}}, \bibnamefont{and}
  \bibinfo{author}{\bibfnamefont{S.}~\bibnamefont{Das~Sarma}},
  \bibinfo{journal}{Rev. Mod. Phys.} \textbf{\bibinfo{volume}{80}},
  \bibinfo{pages}{1083} (\bibinfo{year}{2008}).

\bibitem[{\citenamefont{Wen}(2004)}]{Wen2004B}
\bibinfo{author}{\bibfnamefont{X.-G.} \bibnamefont{Wen}},
  \emph{\bibinfo{title}{Quantum Field Theory Of Many-body Systems: From The
  Origin Of Sound To An Origin Of Light And Electrons}}
  (\bibinfo{publisher}{Oxford University Press, New York},
  \bibinfo{year}{2004}).

\bibitem[{\citenamefont{Halperin}(1983)}]{Halperin1983}
\bibinfo{author}{\bibfnamefont{B.~I.} \bibnamefont{Halperin}},
  \bibinfo{journal}{Helv. Phys. Acta} \textbf{\bibinfo{volume}{56}},
  \bibinfo{pages}{75} (\bibinfo{year}{1983}).

\bibitem[{\citenamefont{Levin and Stern}(2012)}]{Levin2012a}
\bibinfo{author}{\bibfnamefont{M.}~\bibnamefont{Levin}} \bibnamefont{and}
  \bibinfo{author}{\bibfnamefont{A.}~\bibnamefont{Stern}},
  \bibinfo{journal}{Phys. Rev. B} \textbf{\bibinfo{volume}{86}},
  \bibinfo{pages}{115131} (\bibinfo{year}{2012}).

\bibitem[{\citenamefont{Lu and Vishwanath}(2012)}]{Lu2012a}
\bibinfo{author}{\bibfnamefont{Y.-M.} \bibnamefont{Lu}} \bibnamefont{and}
  \bibinfo{author}{\bibfnamefont{A.}~\bibnamefont{Vishwanath}},
  \bibinfo{journal}{Phys. Rev. B} \textbf{\bibinfo{volume}{86}},
  \bibinfo{pages}{125119} (\bibinfo{year}{2012}).

\bibitem{sup} For details see supplemental materials.

\bibitem[{\citenamefont{Wen}(1995)}]{Wen1995}
\bibinfo{author}{\bibfnamefont{X.-G.} \bibnamefont{Wen}},
  \bibinfo{journal}{Advances in Physics} \textbf{\bibinfo{volume}{44}},
  \bibinfo{pages}{405} (\bibinfo{year}{1995}), ISSN \bibinfo{issn}{0001-8732}.

\bibitem[{\citenamefont{Hermele et~al.}(2009)\citenamefont{Hermele, Gurarie,
  and Rey}}]{Hermele2009}
\bibinfo{author}{\bibfnamefont{M.}~\bibnamefont{Hermele}},
  \bibinfo{author}{\bibfnamefont{V.}~\bibnamefont{Gurarie}}, \bibnamefont{and}
  \bibinfo{author}{\bibfnamefont{A.~M.} \bibnamefont{Rey}},
  \bibinfo{journal}{Phys. Rev. Lett.} \textbf{\bibinfo{volume}{103}},
  \bibinfo{pages}{135301} (\bibinfo{year}{2009}).

\bibitem[{\citenamefont{Schroeter et~al.}(2007)\citenamefont{Schroeter, Kapit,
  Thomale, and Greiter}}]{Schroeter2007}
\bibinfo{author}{\bibfnamefont{D.~F.} \bibnamefont{Schroeter}},
  \bibinfo{author}{\bibfnamefont{E.}~\bibnamefont{Kapit}},
  \bibinfo{author}{\bibfnamefont{R.}~\bibnamefont{Thomale}}, \bibnamefont{and}
  \bibinfo{author}{\bibfnamefont{M.}~\bibnamefont{Greiter}},
  \bibinfo{journal}{Phys. Rev. Lett.} \textbf{\bibinfo{volume}{99}},
  \bibinfo{pages}{097202} (\bibinfo{year}{2007}).

\bibitem[{\citenamefont{{Lu} and {Lee}}(2012)}]{Lu2012c}
\bibinfo{author}{\bibfnamefont{Y.-M.} \bibnamefont{{Lu}}} \bibnamefont{and}
  \bibinfo{author}{\bibfnamefont{D.-H.} \bibnamefont{{Lee}}},
  \bibinfo{journal}{ArXiv e-prints 1212.0863}  (\bibinfo{year}{2012}),
  \eprint{1212.0863}.

\bibitem[{\citenamefont{Kitaev}(2006)}]{Kitaev2006}
\bibinfo{author}{\bibfnamefont{A.}~\bibnamefont{Kitaev}},
  \bibinfo{journal}{Annals of Physics} \textbf{\bibinfo{volume}{321}},
  \bibinfo{pages}{2} (\bibinfo{year}{2006}), ISSN \bibinfo{issn}{0003-4916}.

\bibitem[{\citenamefont{Lu and Vishwanath}(2013)}]{Lu2013}
\bibinfo{author}{\bibfnamefont{Y.-M.} \bibnamefont{Lu}} \bibnamefont{and}
  \bibinfo{author}{\bibfnamefont{A.}~\bibnamefont{Vishwanath}},
  \bibinfo{journal}{ArXiv e-prints 1302.2634}  (\bibinfo{year}{2013}).

\bibitem[{\citenamefont{Miranda and Morrison}(2009)}]{Miranda2009}
\bibinfo{author}{\bibfnamefont{R.}~\bibnamefont{Miranda}} \bibnamefont{and}
  \bibinfo{author}{\bibfnamefont{D.~R.} \bibnamefont{Morrison}}, \bibinfo{title}{Embeddings of Integral Quadratic Forms},
  (\bibinfo{year}{2009}).

\bibitem[{\citenamefont{Essin and Hermele}(2013)}]{Essin2013}
\bibinfo{author}{\bibfnamefont{A.~M.} \bibnamefont{Essin}} \bibnamefont{and}
  \bibinfo{author}{\bibfnamefont{M.}~\bibnamefont{Hermele}},
  \bibinfo{journal}{Phys. Rev. B} \textbf{\bibinfo{volume}{87}},
  \bibinfo{pages}{104406} (\bibinfo{year}{2013}).

\bibitem[{\citenamefont{Mesaros and Ran}(2013)}]{Mesaros2013}
\bibinfo{author}{\bibfnamefont{A.}~\bibnamefont{Mesaros}} \bibnamefont{and}
  \bibinfo{author}{\bibfnamefont{Y.}~\bibnamefont{Ran}},
  \bibinfo{journal}{Phys. Rev. B} \textbf{\bibinfo{volume}{87}},
  \bibinfo{pages}{155115} (\bibinfo{year}{2013}).

\bibitem[{\citenamefont{Hung and Wen}(2013)}]{Hung2013}
\bibinfo{author}{\bibfnamefont{L.-Y.} \bibnamefont{Hung}} \bibnamefont{and}
  \bibinfo{author}{\bibfnamefont{X.-G.} \bibnamefont{Wen}},
  \bibinfo{journal}{Phys. Rev. B} \textbf{\bibinfo{volume}{87}},
  \bibinfo{pages}{165107} (\bibinfo{year}{2013}).

\bibitem[{\citenamefont{Hung and Wan}(2013)}]{Hung2013a}
\bibinfo{author}{\bibfnamefont{L.-Y.} \bibnamefont{Hung}} \bibnamefont{and}
  \bibinfo{author}{\bibfnamefont{Y.}~\bibnamefont{Wan}},
  \bibinfo{journal}{Phys. Rev. B} \textbf{\bibinfo{volume}{87}},
  \bibinfo{pages}{195103} (\bibinfo{year}{2013}).

\bibitem[{\citenamefont{{Cano} et~al.}(2013)\citenamefont{{Cano}, {Cheng},
  {Mulligan}, {Nayak}, {Plamadeala}, and {Yard}}}]{Cano2013}
\bibinfo{author}{\bibfnamefont{J.}~\bibnamefont{{Cano}}},
  \bibinfo{author}{\bibfnamefont{M.}~\bibnamefont{{Cheng}}},
  \bibinfo{author}{\bibfnamefont{M.}~\bibnamefont{{Mulligan}}},
  \bibinfo{author}{\bibfnamefont{C.}~\bibnamefont{{Nayak}}},
  \bibinfo{author}{\bibfnamefont{E.}~\bibnamefont{{Plamadeala}}},
  \bibnamefont{and} \bibinfo{author}{\bibfnamefont{J.}~\bibnamefont{{Yard}}},
  \bibinfo{journal}{ArXiv e-prints}  (\bibinfo{year}{2013}),
  \eprint{1310.5708}.

\bibitem[{\citenamefont{Read}(1990)}]{Read1990}
\bibinfo{author}{\bibfnamefont{N.}~\bibnamefont{Read}}, \bibinfo{journal}{Phys.
  Rev. Lett.} \textbf{\bibinfo{volume}{65}}, \bibinfo{pages}{1502}
  (\bibinfo{year}{1990}).

\bibitem[{\citenamefont{Frohlich and Zee}(1991)}]{Frohlich1991}
\bibinfo{author}{\bibfnamefont{J.}~\bibnamefont{Frohlich}} \bibnamefont{and}
  \bibinfo{author}{\bibfnamefont{A.}~\bibnamefont{Zee}},
  \bibinfo{journal}{Nuclear Physics B} \textbf{\bibinfo{volume}{364}},
  \bibinfo{pages}{517} (\bibinfo{year}{1991}), ISSN \bibinfo{issn}{0550-3213}.

\bibitem[{\citenamefont{Wen and Zee}(1992)}]{Wen1992}
\bibinfo{author}{\bibfnamefont{X.~G.} \bibnamefont{Wen}} \bibnamefont{and}
  \bibinfo{author}{\bibfnamefont{A.}~\bibnamefont{Zee}},
  \bibinfo{journal}{Phys. Rev. B} \textbf{\bibinfo{volume}{46}},
  \bibinfo{pages}{2290} (\bibinfo{year}{1992}).

\end{thebibliography}

\end{document}